# Beam optimization of RFQ and SFRFQ combined accelerator at Peking University

**Minglei Kang, Yuanrong Lu #, Zhi Wang , Kun Zhu, Xueqing Yan, Shuli Gao, Shixiang Peng, Zhiyu Guo, Jiaxun Fang , Jiaer Chen**

State key laboratory of nuclear physics and technology, Peking University，Beijing100871, China

*E-mail*: mlkang@pku.edu.cn

The Peking University Integral Split Ring Radio Frequency Quadrupole(ISR RFQ) accelerator was constructed in 1999 with a high duty factor 16.7% and repetition frequency 166Hz, and it was able to accelerate $N^+$, $O^+$, $O^-$, $C^+$ and $He^+$ from 1.6kev/u to 65keV/u[1]. It was later upgraded as an injector of the Separated Function RFQ (SFRFQ) [2, 3]. The experiments indicated that the maximum accelerated $O^+$ beam current could exceed 3.2mA with energy 1.03MeV and an energy spread (FWHM) 3.1%. Then the beam transports through a 1m-long magnetic triplet to the entrance of SFRFQ and is finally accelerated to 1.64MeV [4]. The beam conditioning of RFQ were carefully optimized to satisfy the requirements of the SFRFQ. The combined accelerator eventually can deliver 0.53mA $O^+$ beam with energy 1.65MeV, which has sufficiently demonstrated the feasibility of the SFRFQ structure.

## 1. Introduction

Separated Function Radio Frequency Quadrupole (SFRFQ) accelerator is a new structure that uses gaps and quadrupole electric field to accelerate and focus particles in the longitudinal and transverse direction. It promises higher accelerating efficiency than the conventional RFQ. In order to demonstrate its feasibility a full prototype cavity was constructed as a post accelerator of ISR RFQ.

The combined accelerator is a complex system including two different accelerating structures: RFQ and SFRFQ. The performance of RFQ accelerator is one of the crucial factors to achieve the SFRFQ dynamics design. The dynamics design of RFQ was given by PARMTEQ, which is a classical Four-Section Procedure from the Los Alamos National Laboratory (LANL) [5]. If the beam current was less than 1mA, the transmission was 86% [6]. After upgrading of the ECR ion source, the extracted $O^+$ beam exceeded 6mA, the

---
# Corresponding author: yrlu@pku.edu.cn
Work supported by NSFC, contract No.10905003, China Postdoctoral Science Foundation

transmission and emittance status would be much different from the previous low current beam. So the beam optimization experiments were necessary to search for the best operating parameters.

In this paper, section 2 is a description of the beam line setup and rf system; section 3 presents the process of the beam optimization of ECR ion source and RFQ, section 4 is the results of beam commissioning of the combined accelerator; and the conclusion is given in section 5.

## 2. The beam line components and rf system

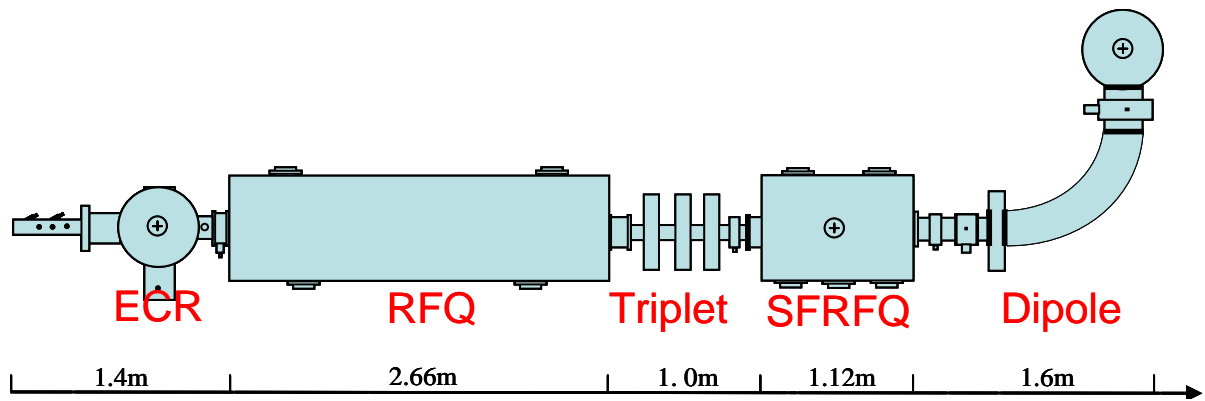

Fig 1. Layout of RFQ and SFRFQ combined accelerator system

The beam line consists of five parts as shown in Fig 1: the 2.45GHz permanent magnet electron cyclotron resonace(ECR) ion source and Low energy beam transport (LEBT) system , a 2.66m-long traditional RFQ accelerator, a magnetic triplet transverse matching section, a 1.12m-long SFRFQ accelerator, and energy analyzing magnet(AM). There are 4 faraday cups installed in the beam line,  each was located at the entrance and exit of the RFQ and SFRFQ.

Tbale 1

| Parameters of ECR Ion sources | |
|---|---|
| Frequency | 2.45GHz |
| Ions | $N^+, O^+,$ |
| Extraction voltage | 20~28kV |
| Duty factor | 1/6 |
| Emittance (N) | <0.2 mm mrad |

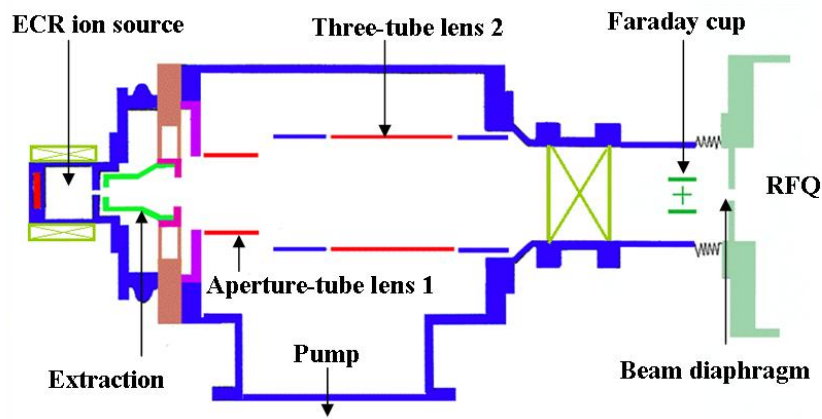

Fig 2 ECR ion source and Low energy beam transmission (LEBT)

The parameters of ECR ion source are presented in table 1. It operates at 2.45GHz in a pulse mode, and the pulse duration is 1ms and the repetition is 166Hz. It adopted a dipole extraction system, the emission aperture radius on the plasma electrode is 2.5mm [7], and the accelerating gap was 2mm. There are two electrostatic lenses with diameters 80mm and 120mm (lens 1 and 2) be used in the LEBT section to focus the extracted beams. The first lens is an aperture-tube lens ($U_1$), the second one is a Three-tube Lens, and the length of two ground tubes and the middle one ($U_2$) are 50mm and 120mm, respectively. A diaphragm with aperture radius 7.5mm been installed between the first faraday cup and RFQ to scrape off large divergence angle particles, and experiments indicated that only about 2/3 beam that received by the faraday cup can transport into the RFQ cavity[6]

The RFQ receives $O^+$ beam from ECR ion source, then accelerates the beam to 1.03MeV. The first prototype SFRFQ will accelerate the bunched beam from the RFQ to 1.64MeV. A magnet triplet is used as a matching section connecting RFQ and SFRFQ. The energy will be analyzed by a dipole magnet at the end of the beam line.

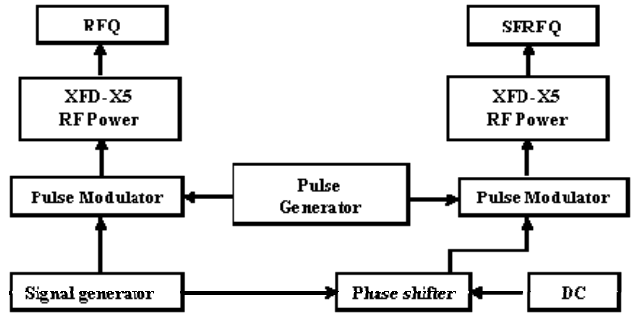

Fig 3.The schematic of rf power feeding

The RFQ and SFRFQ accelerators are fed by two XFD-X5 rf power sources, respectively, as illustrated in Fig 3. Each power source can deliver 30kW on CW mode and maximum 45kW on pulse mode with 1/6 duty circle [8]. The rf phase difference between the RFQ and SFRFQ cavity was changed by a phase shifter in the rf circuit of SFRFQ and its range is 0 to 360 degree.

## 3. Beam optimization of ISR RFQ

The fraction of $O^+$ from the plasma chamber was about 80% [6], and then the beam transported through the LEBT to the entrance of RFQ. For a given extraction voltage, the focusing voltage $U_1$ and $U_2$ can be optimized to get maximum beam current at the entrance of RFQ. Obviously, when the gas flow, the microwave power feeding etc. were kept constant, the beam current increased with extraction voltage, as illustrated in Fig4 (a), the extracted maximum current exceeded 7.5mA. Correspondingly, the optimal focusing voltage $U_1$ and $U_2$ also increased with the extraction voltage (see Fig 4 (b)).

The discharged pressure of the ion source can start from $1.0*10^{-4}$Pa[7], while the optimal working pressure should be set larger than $2.5*10^{-4}$Pa, and the microwave power feeding shouldn't be less than 227W in average and 1366W in peak.

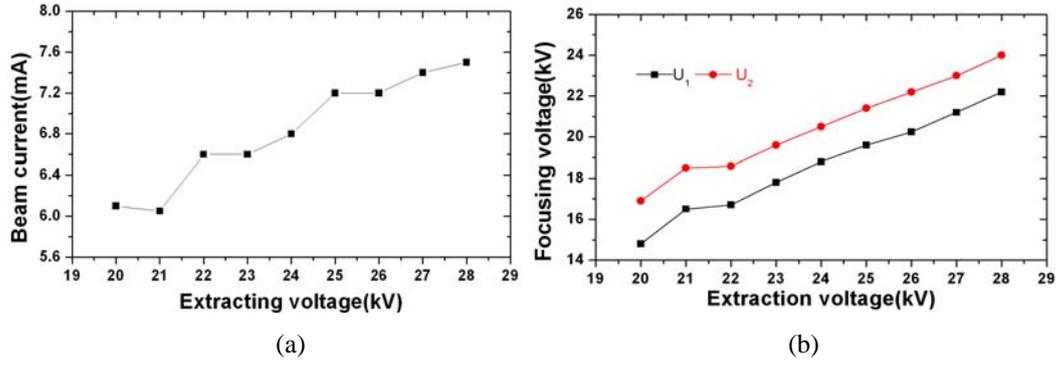

(a)                                         (b)

Fig 4 Extracting voltage against beam current of Ion source (a) and focusing voltage (b)

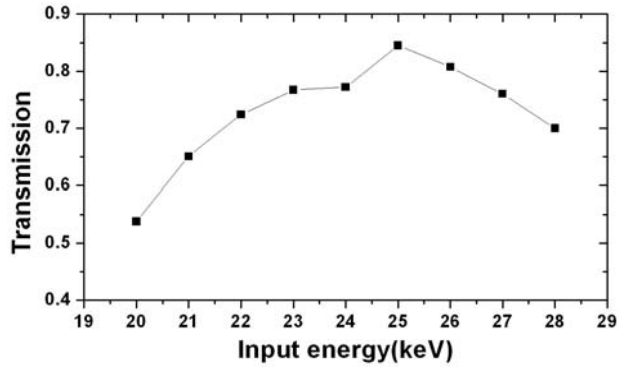

Fig 5.  Input energy vs. RFQ transmission

    The original designed input energy of the RFQ is 22keV, while the experiment shows that the RFQ has maximum transmission when the input energy is 25keV (see Fig 5). This can be explained as following .There are two kinds of particles losses in the RFQ, one is induced by transverse reasons named T-type and the other is induced by longitudinal reasons named L-type [9]. The T-type occurs when beam is poorly matched or space charge effect is under estimated. The designed peak current of ISR RFQ accelerator was 1mA, after upgrading of ECR ion source the accelerated beam reached 3.2mA, the space charge effect could make the transverse emittance grow greatly. So improving the input energy of beam is a good choice both to reduce space charge forces and T-type losses. The transmission increases with the improvement of input energy, but if the input energy exceeds a certain value, which will induce the L-type losses. Some particles can not stay in bucket of stable oscillations around the synchronous phase; they will slip toward the following bunch. Some of them couldn't catch stable phases and get decelerated. The L-type eventually converts to T-type and particles get lost at radial direction. So when improving the input energy from 22keV to 27keV, firstly space charge was reduced, when the energy exceeds 25keV the L-type losses would become serious and transmission will decline.

    The power feeding of the RFQ was also optimized to get high beam transmission. The input power is cost in two ways, one is the heat losses because of the resistance of cavity, and the other one is carried away by particles. If the feeding power is not enough, the designed inter-vane voltage can't be reached, on the contrary if the power is too high, the inter-vane voltage will exceed the designed value, and both of these two cases will bring L-type losses. We changed the input feeding power from 28kW to 38kW with duty circle 1/6, and then found there was an optimal rf power feeding 35kW as shown in Fig 6.

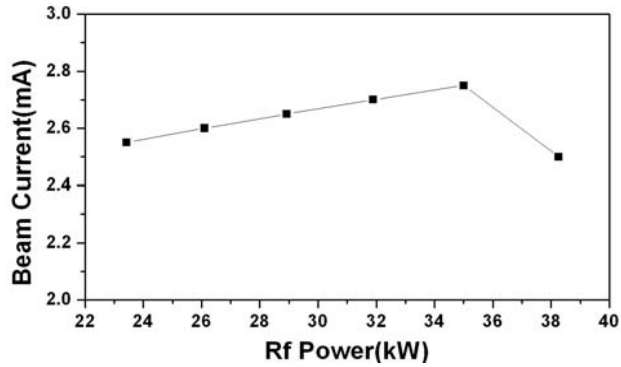

Fig 6 Power feeding of RFQ vs beam current (extraction voltage was 22.5kV, and the focusing voltage lens 1 and lens 2 were 17.4 and 19.3kV, respectively)

Eventually the best operating parameters of the RFQ were: extraction voltage was set at 25kV, focusing voltage $U_1$ and $U_2$ were 19.3 and 21.4kV, the feeding power was 35kW with 1/6 duty circle. The maximum accelerated $O^+$ beam reached 3.2mA with a transmission about 84%. The energy spectrum was measured and the peak energy was 1.03MeV with a spread 3.1%.

## 4. The beam commissioning of combined accelerator system

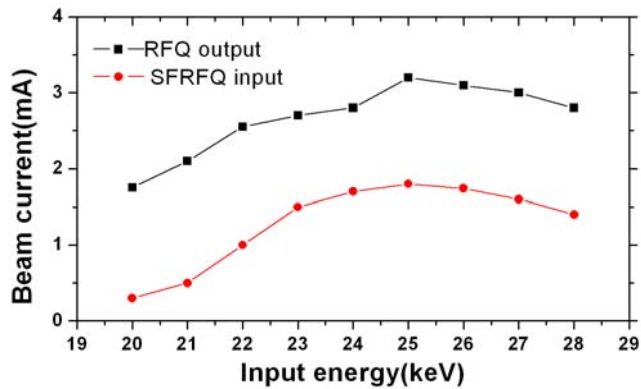

Fig 7. The input energy of RFQ vs output beam current of RFQ and input of SFRFQ

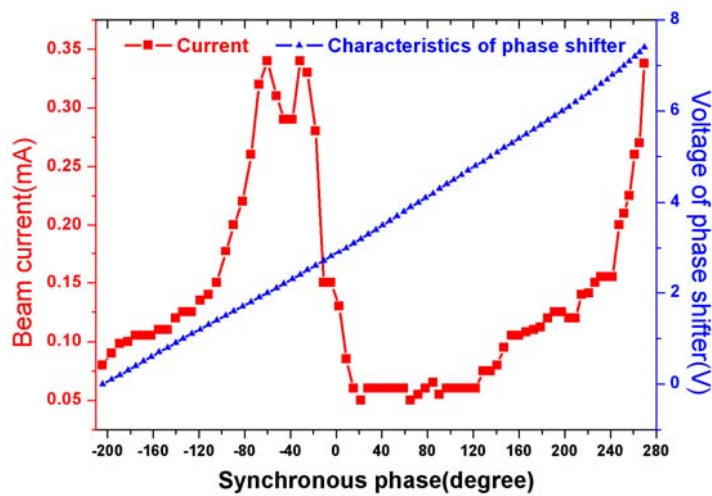

Fig 8 Synchronous phase vs beam current of SFRFQ, and the characteristics of phase shifter (the beam current is a function of synchronous phase and it can be repeated after 360 degree)

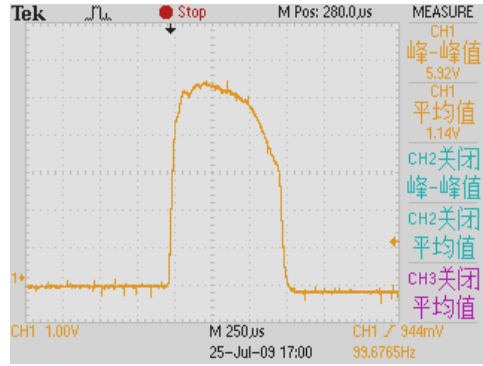

Fig 9 The accelerated beam by SFRFQ was recorded by a faraday cup with a 10koms sample resistor

The maximum beam current at the exit of RFQ is 3.2mA, but first accelerated beam current by SFRFQ was only 0.2mA. A lot of particles hit the wall and got lost. In order to figure out the reason of low transmission, the transverse emittance was measured at the end of the RFQ accelerator. The measured $e_x$=0.25mmmrad and $e_y$=0.48mmmrad, which are much larger than the designed 0.1mmrad by PARMTEQM. The RFQ was designed with peak current 1mA, so the space charge effect increases greatly after the upgrading of ECR ion source. Therefore a diaphragm with radius 5mm was installed before the entrance of SFRFQ, first it could decrease the beam intensity make sure the electrode of SFRFQ not be damaged, second it could scrape off large angle particles and get a relative precise transmission. As presented in Fig 7, the input beam current of SFRFQ was much smaller than the output of RFQ. After transporting through the 1m-long magnetic triplet and the diaphragm to the entrance of SFRFQ there was only 1.8mA $O^+$ beam been measured.

The SFRFQ uses rf quadrupole electric fields and gaps to focus and accelerate beam in transverse and longitudinal direction, so it is similar to normal DTL structure in z-axis [10]. It receives bunched beam form RFQ, and the synchronous phase of the bunch center should agree with the designed value. The synchronous phase can be set by the phase shifter in the rf circuit. When keeping the parameters constant, such as extraction voltage, focusing $U_1$ and $U_2$, gas pressure, and power feeding RFQ and SFRFQ etc, and from changing the rf phase by turning the control voltage of phase shifter, we found that the beam current of SFRFQ was a function of synchronous phase in the range 0 to 360 degree. As illustrated in Fig 8. The square spots are current. The triangular spots in Fig 8 are the characteristic of phase shifter; the phase curve had linear relation in the voltage range from 0 to 7.5V. Finally, the rf feeding of SFRFQ accelerator was 18.7kW with 1/6 duty circle, the measured beam current was 0.53mA as shown in Fig 9. The energy spectrum indicated that the SFRFQ had accelerated $O^+$ from 1.03MeV to 1.65MeV with an energy spread 3.1%, which agrees well with the dynamics design. [11]

## 5. Conclusion

The PKU ISR RFQ was constructed in 1999 and when the input beam current was lower than 1mA and its transmission efficiency was 86%. Afterwards the ECR was upgraded and the extracted beam exceeding 6mA and the RFQ's transmission was about 84%. After parameter optimization the RFQ can deliver 3.2mA $O^+$ beam from 22keV to 1.03MeV, but the transverse emittance increased greatly. The mismatch of beam between the two accelerators was a biggest barrier when improving the transmission of SFRFQ. Finally the SFRFQ could accelerate O$^+$ beam with 0.53mA. Although the transmission is low, it has sufficiently verified the feasibility of the SFRFQ structure.

**Acknowledgment:**

The author thanks the help from Zhou quanfeng, Zhang Meng, Ren Haitao and Lu Pengnan who are PhD students of Peking University.

**Reference**

[1] Y .R . Lu, J.E Chen, J.X. Fang , et al. *Investigation of high duty factor ISR RFQ-1000. Nucl. Instr. and Meth A, 2003, 515: 394-401.*
[2] J.E Chen, J.X. Fang, W.G. Li, et al. *Tentative idea on accelerating structures of SFRFQ. Progress in Natural Science, 2002, 12(1):22-28*
[3] X.Q. Yan, J.E. Chen, J.X. Fang, et al., *Nucl. Instr. and Meth. A, 2005,* **539**, *606–612*
[4] J.E Chen, Z.Y. Guo, Y .R . Lu, et al. *R&D status of SFRFQ accelerator system. Chinese Physics C, 2008, 32:231-233*
[5]K. R. Crandall, R. H. Stokes, and T. P. Wangler, in Proceedings of the 1979 Linac Conference, Montauk, NY (BNL Report No. BNL-51134, 1979), p. 205.
[6] M .Zhang, Y. R. Lu, S. X. Peng et al. *Upgrading of a heavy ion 1MeV ISR RFQ accelerator. Chinese physics C, 2008, 32: 220*
[7] S. X. Peng, M. Zhang，Z. Z. Song, et al. *Experimental results of an electron cyclotron resonance oxygen source and a low energy beam transport system for 1 MeV integral split ring radio frequency quadruple accelerator upgrade project. Rev. Sci. Instrum, 2008,79, 02B706*
[8] Y. R. Lu, J. E. Chen, K. Zhu, et al., *An intermediate structure SFRFQ between RFQ and DTL. Proceedings of the 2008 LINAC conference, Victoria, BC, Canada, 2008:118-120*
[9] C. Zhang, Z. Y. Guo, A. Schempp, et al., *Low-beam-loss design of a compact high-current deuteron radio frequency quadrupole accelerator. PHYSICAL REVIEW SPECIAL TOPICS- ACCELERATORS AND BEAMS, 2004 ,7, 100101*
[10] Z. Wang, J.E. Chen, Y.R. Lu, et al., *Nucl. Instr. and Meth. A , 2007,***572**, *596*
[11] M. L. Kang，Y. R. Lu， J. E. Chen，et al., *in Proceeding of 2010 Chinese Particle Accelerator Conference, Beijing, China, 2010*